\documentclass[aps,twocolumn,superscriptaddress]{revtex4}

\usepackage{amsmath,amssymb}
\usepackage{epsfig}
\usepackage{graphicx}
\begin{document}

\title{
Gravity and large black holes in Randall-Sundrum II braneworlds
}

\author{Pau Figueras}
\affiliation{DAMTP, Centre for Mathematical Sciences, University of Cambridge, Wilberforce Road, Cambridge CB3 0WA, U.K.}

\author{Toby Wiseman}
\affiliation{Theoretical Physics Group, Blackett Laboratory, Imperial College, London SW7 2AZ, U.K.}

\date{May 2011}

\begin{abstract}

We show how to construct low energy solutions to the Randall Sundrum II (RSII) model using an associated $AdS_5$-$CFT_4$ problem. The RSII solution is given in terms of a perturbation of the $AdS_5$-$CFT_4$ solution, with the perturbation parameter being the radius of curvature of the brane metric compared to the AdS length $\ell$. The brane metric is then a specific perturbation of the $AdS_5$-$CFT_4$ boundary metric.
For low curvatures the RSII solution reproduces 4$d$ GR on the brane. The leading correction is from local higher derivative curvature terms. The subleading correction is derived from similar terms and also the dual CFT stress tensor.
Recently $AdS_5$-$CFT_4$ solutions with 4$d$ Schwarzschild boundary metric were numerically constructed. We modify the boundary conditions to introduce the RSII brane, and use elliptic numerical methods to solve the resulting boundary value problem.
We construct large RSII static black holes with radius up to $\sim 20 \ell$. For large radius the RSII solutions are indeed close to the associated $AdS_5$-$CFT_4$ solution.
In this case the local curvature corrections vanish, and we confirm the leading correction is given by the $AdS_5$-$CFT_4$ solution stress tensor. 
We also follow the black holes to small radius $\ll\ell $, where as expected they transition to a 5$d$  behaviour. 
Our numerical solutions indicate the RSII black holes are dynamically stable
for axisymmetric perturbations
for all radii.
\end{abstract}


\maketitle

\section{Introduction}

The single brane RSII model \cite{RSI,RSII} is remarkable in that it is claimed to yield 4$d$ low energy physics for brane observers even though the 5$d$ geometry is not compact. 
Using arguments from $AdS$-$CFT$ it has been claimed that the low energy behaviour of this model for a brane observer is equivalent to 4$d$ gravity coupled to a conformal field theory (CFT) \cite{Verlinde:1999fy,Gubser:1999vj,Hawking:2000kj,Duff:2000mt,Giddings:2000mu,deHaro:2000wj}\footnote{This was first discussed in unpublished remarks by  Juan Maldacena and by Edward Witten.}. Following from this, a remarkable conjecture has been made in \cite{Tanaka:2002rb,Emparan:2002px} that static black holes can not exist in RSII for radius much greater than the AdS length, $\ell$. 
If true this is an important phenomenological result, allowing constraints on the existence of extra dimensions derived from astrophysical black holes rather than tests of Newton's law \cite{Emparan:2002jp}. However the conjecture is based on applying free field theory intuition to the CFT, which is strongly coupled. It has been argued that such extrapolation may not be justified \cite{Fitzpatrick:2006cd, Gregory:2008br}. 

Ultimately the existence of RSII black hole solutions reduces to existence of solutions to the non-linear coupled PDEs of the Einstein equations. Using  the numerical methods of \cite{Wiseman:2001xt,Wiseman:2002zc},
 black holes in 5$d$ RSII with radius up to $\sim 0.2 \ell$, and for 6$d$ up to $\sim 2.0 \ell$  were constructed in \cite{Kudoh:2003xz,Kudoh:2003vg,Kudoh:2004kf}. However, using the same methods it has subsequently been argued that even very small RSII static black holes do not exist \cite{Yoshino:2008rx,Kleihaus:2011yq}. Other numerical work includes \cite{Chamblin:2000ra,Casadio:2002uv} and perturbative construction of small black holes was done in \cite{Karasik:2003tx,Karasik:2004wk}. The near horizon geometry of extremal RSII black holes has been determined \cite{Kaus:2009cg}, although extremal solutions are thought to evade the non-existence conjecture.
 
In this letter we  firstly will make precise the claim that low energy physics on the brane is described by gravity coupled to a CFT. 
We shall explicitly show how to construct low curvature solutions to RSII, including matter on the brane, from an associated $AdS_5$-$CFT_4$ problem, where the boundary metric is given by a particular perturbation of the brane metric. 
An $AdS_5$-$CFT_4$ solution with Schwarzschild boundary metric has recently been numerically constructed by us and Lucietti \cite{PFLuciettiTW} and in the second half of the letter we shall report on work where we modify the numerical construction used there to compute the RSII black hole solutions for both large and small radii. For large radius the solutions are close to this $AdS_5$-$CFT_4$ solution. The details of this numerical construction and results will be discussed in a forthcoming paper \cite{PFTW}, and in this letter we give an overview of the methodology and highlight the salient results.

\section{Low curvature RSII solutions from $AdS_5$-$CFT_4$}
\label{sec:RS2}

In this section we will follow \cite{deHaro:2000wj} although we note the emphasis is subtly different. Our aim is not to derive an effective 4$d$ description of gravity on the brane as done in \cite{deHaro:2000wj}, but rather to explicitly demonstrate the relation between solutions in AdS/CFT and corresponding ones in RSII.

Consider a solution to $AdS_5$-$CFT_4$ with boundary metric $g^{(0)}_{\mu\nu}$. The 5$d$ metric $g_{AB}$ obeying $R_{AB} = - \frac{4}{\ell^2}\, g_{AB} $ can be written as,
\begin{eqnarray}
ds^2 = g_{AB} dx^A dx^B = \frac{\ell^2}{z^2} \left( dz^2 + \tilde{g}_{\mu\nu}(z,x) dx^\mu dx^\nu \right)
\label{eqn:FGmetric}
\end{eqnarray}
near the conformal boundary, $z = 0$, where the Fefferman-Graham expansion dictates that, 
\begin{eqnarray}
 \tilde{g}_{\mu\nu}(z,x)  =  g^{(0)}_{\mu\nu}(x) + z^2 \left( R^{(0)}_{\mu\nu}(x) - \frac{1}{4} g^{(0)}_{\mu\nu}(x) R^{(0)}(x) \right) \nonumber \\
 + z^4 \left( g^{(4)}_{\mu\nu}(x) + t_{\mu\nu}(x) \right) + 2 z^4 \log{z} \, h^{(4)}_{\mu\nu}(x) + O(z^6)
 \end{eqnarray}
 where the expressions for $g^{(4)}$ and $h^{(4)}$ can be found in \cite{deHaro:2000xn}.
Here $g^{(0)}_{\mu\nu}(x)$ and $t_{\mu\nu}(x)$ are the two constants of integration for the bulk equations which are second order in $z$. 
The constraint equations for this radial evolution imply $\nabla^{(0)}_\mu t^{\mu\nu} = 0$ and $t = \frac{1}{16} \left( R^{(0)}_{\alpha\beta} R^{(0)\alpha\beta} - \frac{1}{3} ( R^{(0)} )^2 \right)$, and $t_{\mu\nu}$ gives the vev of the $CFT_4$ stress tensor as, $\langle T^{CFT}_{\mu\nu} \rangle = t_{\mu\nu} / (4 \pi \ell G_5)$. 

We assume that for some boundary metric 
$g^{(0)}_{\mu\nu} = {g}_{\mu\nu}$
a solution exists for boundary conditions in the IR of the geometry such that the metric tends to the Poincare horizon of AdS. 
We further assume that solutions exist for regular perturbations of the boundary metric $g^{(0)}_{\mu\nu} = {g}_{\mu\nu} + \epsilon^2 h_{\mu\nu}$ in some finite neighbourhood of $\epsilon = 0$, so that,
$t_{\mu\nu}[ {g} + \epsilon^2 h ] = t_{\mu\nu}[ {g} ] + O(\epsilon^2)$.
   
From this $AdS_5$-$CFT_4$ solution we will construct an RSII solution in the limit where brane curvatures are small compared to the curvature of the bulk $AdS_5$. We take two copies of the solution above restricted to $z \ge \epsilon$ and glue them together on their common boundary. We then identify the two halves under a $\mathbb{Z}_2$ action which leaves the orbifold plane $z = \epsilon$, the RSII brane with induced metric $\gamma_{\mu\nu}$, 
fixed. The Israel conditions determine the matter on the brane to have stress tensor,
\begin{eqnarray}
\label{eq:T}
8 \pi G_4 T^{brane}_{\mu\nu} & = & \frac{2}{\ell} \left( K_{\mu\nu} - K \gamma_{\mu\nu} + \frac{3}{\ell} \gamma_{\mu\nu} \right)
 \end{eqnarray}
  where $K_{\mu\nu} = - \frac{1}{2} \frac{z}{\ell} \partial_z\big(\frac{\ell^2}{z^2} \,\tilde{g}_{\mu\nu}(z,x)\big)$ is the extrinsic curvature of the $z = \epsilon$ surface.
  
We begin the construction by choosing the perturbation $h_{\mu\nu}$ so that $\gamma_{\mu\nu} = \frac{\ell^2}{\epsilon^2} g_{\mu\nu}$, and then,
\begin{eqnarray}
\label{eq:boundary}
g^{(0)}_{\mu\nu} &=& {g}_{\mu\nu} + \frac{\epsilon^2}{2} \left( {R}_{\mu\nu} - \frac{1}{6} {g}_{\mu\nu} {R} \right) + O( \epsilon^4 \log{\epsilon} ) \, .
\end{eqnarray}
It is convenient to work with the rescaled brane metric $g_{\mu\nu}$, rather than $\gamma_{\mu\nu}$ since we are interested in the limit $\epsilon \rightarrow 0$.

Computing the brane matter stress tensor from \eqref{eq:T} in terms of the rescaled brane metric $g_{\mu\nu}$ gives the `Einstein equation on the brane' 
derived in \cite{deHaro:2000wj}:
\begin{eqnarray}
&& G_{\mu\nu} - 8 \pi G_4 T^{brane}_{\mu\nu} =   \epsilon^2 \log{\epsilon}\,  {b}_{\mu\nu}[g] \\
& & \qquad +  \epsilon^2 \left( 16 \pi G_4  \langle T^{CFT}_{\mu\nu}[g] \rangle +  {a}_{\mu\nu}[g]  \right) + O(\epsilon^4 \log \epsilon) \nonumber
 \end{eqnarray} 
 where (the separately conserved) tensors $a_{\mu\nu}$,  $b_{\mu\nu}$ are,
 \begin{eqnarray}
a_{\mu\nu}[{g}] & \equiv & -\frac{1}{4}{\nabla}^2{R}_{\mu\nu} + \frac{1}{12}{\nabla}_\mu{\nabla}_\nu{R} + \frac{1}{24} {\nabla}^2 \bar{R}{g}_{\mu\nu} + \frac{1}{6}{R}{R}_{\mu\nu} \nonumber \\
&&  + \frac{1}{8}{R}_{\alpha\beta}{R}^{\alpha\beta}{g}_{\mu\nu} - \frac{1}{24}{R}^2{g}_{\mu\nu} - \frac{1}{2}{R}_{\mu\alpha\nu\beta}{R}^{\alpha\beta} \nonumber \\
b_{\mu\nu}[{g}] & \equiv &  -\frac{1}{2}{\nabla}^2{R}_{\mu\nu} + \frac{1}{6}{\nabla}_\mu{\nabla}_\nu{R} + \frac{1}{12} {\nabla}^2{R}{g}_{\mu\nu} + \frac{1}{3}{R}{R}_{\mu\nu} \nonumber \\
&&  + \frac{1}{4}{R}_{\alpha\beta}{R}^{\alpha\beta}{g}_{\mu\nu} - \frac{1}{12}{R}^2{g}_{\mu\nu} -{R}_{\mu\alpha\nu\beta}{R}^{\alpha\beta} \; .
\end{eqnarray}
The parameter $\epsilon$ controls the curvature scale on the brane relative to the AdS length and $\epsilon \rightarrow 0$ gives the low curvature limit on the brane where we see the usual 4$d$ Einstein equations are recovered. 

Subject to the assumption that the $AdS_5$-$CFT_4$ solution exists for boundary metric \eqref{eq:boundary} we have constructed a braneworld solution with metric $\gamma_{\mu\nu} = \frac{\ell^2}{\epsilon^2} g_{\mu\nu}$ perturbatively in $\epsilon$. Working to higher order in $\epsilon$ one will obtain further local higher curvature terms together with terms involving functional derivatives of the $CFT_4$ stress tensor. We note that we have not assumed $g_{\mu\nu}$ is a metric perturbation of flat space, only that its curvature is everywhere small.

Interestingly the leading correction in $\epsilon$ for 4$d$ Einstein gravity comes from the $O(\epsilon^2 \log \epsilon)$ local four derivative term $b_{\mu\nu}[g]$. 
In the absence of brane matter, $T^{brane}_{\mu\nu} = 0$, then as $g_{\mu\nu}$ is Ricci flat to order $O(\epsilon^0)$, the corrections $a_{\mu\nu}$, $b_{\mu\nu}$ vanish to give,
\begin{eqnarray}
\label{eq:vac}
\delta G_{\mu\nu} = 16 \pi G_4  \langle T^{CFT}_{\mu\nu}[ {g} ] \rangle
\end{eqnarray} 
where 
$G_{\mu\nu}[ g ] = \epsilon^2 \delta G_{\mu\nu} + O(\epsilon^4)$. This form of correction was conjectured by \cite{Emparan:2002px}, and here we have provided a proof of this, although we emphasize that including brane matter, the CFT correction is not the leading one.

\section{5d static RSII black holes}

\noindent \emph{Set up.} In the previous section we have seen how low curvature classical solutions of the RSII model with brane metric 
$\frac{\ell^2}{\epsilon^2} g_{\mu\nu}$
are related to existence of $AdS_5$-$CFT_4$ solutions with boundary metric a perturbation of $g_{\mu\nu}$.
Consider large static vacuum black holes in RSII. Provided there exists a static $AdS_5$-$CFT_4$ solution with 4$d$ Schwarzschild as the boundary metric and which asymptotes to the  Poincare horizon of AdS in the IR, then large black holes in the RSII scenario exist. Furthermore these will be static, since the $AdS_5$-$CFT_4$ solution they derive from
has boundary metric \eqref{eq:boundary} with $g$ being Schwarzschild
 so that $g^{(0)}$ is static (and will be to all orders in $\epsilon$) and the bulk geometry must inherit the isometries of the boundary metric \cite{AndersonChrusciel}.

Such an $AdS_5$-$CFT_4$ solution has recently been found \cite{PFLuciettiTW} using the new numerical approach of \cite{Headrick:2009pv}. In the remainder of this letter we will report on work where we modify this numerical construction to replace the AdS boundary (`UV' end of the geometry) with an RSII brane boundary condition, and solve the resulting elliptic boundary value problem.  The details  will be presented in a longer forthcoming paper \cite{PFTW}.

Following \cite{Headrick:2009pv} we analytically continue our static solution to Euclidean signature, and
consider the solution to the 5$d$ Einstein-DeTurck equations with a negative cosmological constant,
\begin{equation}
\label{eq:DeTurck}
R_{MN}+\frac{4}{\ell^2}\,g_{MN}-\nabla_{(M}\xi_{N)}=0 
\end{equation}
where $\xi^M=g^{PQ}(\Gamma^M_{PQ}-\bar{ \Gamma}^M_{PQ})$ and $\Gamma^M_{PQ}$ is the connection associated to the metric $g_{AB}$ that we want to determine and $\bar{ \Gamma}^M_{PQ}$ is a connection associated to a fixed reference metric $\bar{g}$. For Euclidean signature the above equation is elliptic and can be solved as a boundary value problem 
for well-posed boundary conditions. 

An important point is that a solution to this Einstein-DeTurck equation need not be Einstein if $\xi^A \ne 0$. In favourable situations one can analytically show that solutions with non-zero $\xi^A$, called `Ricci solitons', cannot exist \cite{PFLuciettiTW}. However, even if they may exist, provided the elliptic problem and boundary conditions are well-posed, solutions should be locally unique. Hence, an Einstein solution cannot be arbitrarily close to a soliton solution \cite{Anderson:2006fk}, and one should easily be able to distinguish the Einstein solutions of interest from solitons.

Following \cite{PFLuciettiTW} we will choose a similar ansatz to that used for the $AdS_5$-$CFT_4$ solution with Schwarzschild boundary, namely,
\begin{align}
\label{eqn:metric}
&ds_5^2 = \frac{\ell^2}{\Delta(r,x)^2} \bigg( r^2 T d\tau^2+ \frac{x^2 g(x) S}{f(r)^2}  d\Omega_{(2)}^2+ \frac{4\,A}{f(r)^4}  dr^2\nonumber \\
& \hspace{2cm} + \frac{4\, B}{f(r)^2g(x)}\,dx^2  + \frac{ 2\,r\,x F }{f(r)^3}\, dr dx   \bigg)\,,\\
 &\Delta(r,x)  =  \frac{(1-r^2)+\tilde\beta(1-x^2)}{ \tilde\beta(1 - r^2) }\,,\nonumber
\end{align}
where $f(r)=1-r^2$ and $g(x)=2-x^2$, and $X=\{T,\,S,\,A,\,B,\,F\}$ are smooth functions (to be determined) which depend on $(r,x)$ only. The (dimensionless) coordinates $(r,x)$ both take values in the range $[0,1]$ and we assume
$T, S > 0$ and that $A B -r^2x^2g(x) F^2/16 > 0$ to ensure that the metric is Euclidean with the correct topology. 
In contrast to the setting in \cite{PFLuciettiTW}, now the function $\Delta(r,x)$ does \textit{not} vanish at $x=1$, so there is no `UV' conformal boundary there.
We choose the reference metric $\bar{g}$ to be the metric \eqref{eqn:metric} with $T = A = B = S = 1$ and $F = 0$.

The boundaries of our domain are the same as in \cite{PFLuciettiTW} (and therefore so are the boundary conditions for the functions $X$ \footnote{Here we have rescaled $\tau\to \ell\,\tilde\beta\,\tau$ so that regularity at the horizon $r=0$ requires $T = 4 A$ there and $\tau\sim \tau+\pi$.}), except that now $x=1$ corresponds to the location of the brane.
Here we impose the vacuum Israel matching conditions (eq.\eqref{eq:T} with the l.h.s.  equal to zero) together with  $\xi_x = 0$ and $F = 0$, which imply mixed Neumann-Dirichlet conditions for the various functions $X$. 
Such boundary conditions have been considered in \cite{PFLuciettiTW} where they were shown to give a regular elliptic system.  Furthermore they imply $\partial_n \xi_r = \frac{2}{\ell} \xi_r$ on the brane (where $\partial_n$ denotes the normal derivative) which is compatible with obtaining an Einstein solution with $\xi = 0$ everywhere.
Note that imposing the Israel vacuum condition and both $\xi_r = 0$ and $\xi_x = 0$ on the brane does not give a regular elliptic system
\cite{Anderson:2006fk,PFLuciettiTW}. We remark that for this negative tension orbifold brane there is no maximum principle argument that rules out 
the existence of a soliton solution. Hence we will have to check explicitly that our solution is Einstein and not a soliton -- indeed we have found no solitons.

Finally we note that our metric \eqref{eqn:metric} has the dimensionless parameter $\tilde\beta$ which determines the inverse temperature as $\beta=4\pi\,\tilde\beta\,\ell$. This effectively controls the size of the black hole relative to the cosmological constant scale.

\bigskip
\noindent \emph{Results.} Two approaches have been proposed  in \cite{Headrick:2009pv} to solve \eqref{eq:DeTurck}. The Ricci flow method works particularly well in finding the $AdS_5$-$CFT_4$ solution in \cite{PFLuciettiTW} since the solution is a stable fixed point of the flow. 
All the RSII black holes we have found have a \emph{single} Euclidean negative mode, and hence the solution is an unstable fixed point of the Ricci flow which makes this method less practical.
For this reason we have used the Newton algorithm to find solutions. We have used two independent codes: one  is based on a pseudospectral collocation approximation in $r$, $x$ (up to $40\times 40$ points), and the other is based on second order finite difference. 
As expected the former gives highly accurate results and the data presented is for this code. The finite difference code gives consistent, but less accurate solutions for the resolutions attainable.

To construct  black holes whose proper radius on the brane, $R_4$, is large compared to $\ell$ (for instance, setting $\tilde\beta=20$ in \eqref{eqn:metric}), we found that using the reference metric $\bar{g}$ as the initial guess was sufficient for Newton's method to converge. Once a large black hole had been obtained we could easily find nearby ones by simply perturbing both the previous solution and the reference metric varying $\tilde\beta$. Using this procedure we have been able to construct braneworld black holes with 
$R_4/\ell\in [0.07,20]$. It should be possible to extend this range increasing the resolution but we have not attempted to do so.

\begin{figure}[h]
\begin{center}
\includegraphics[scale=0.55]{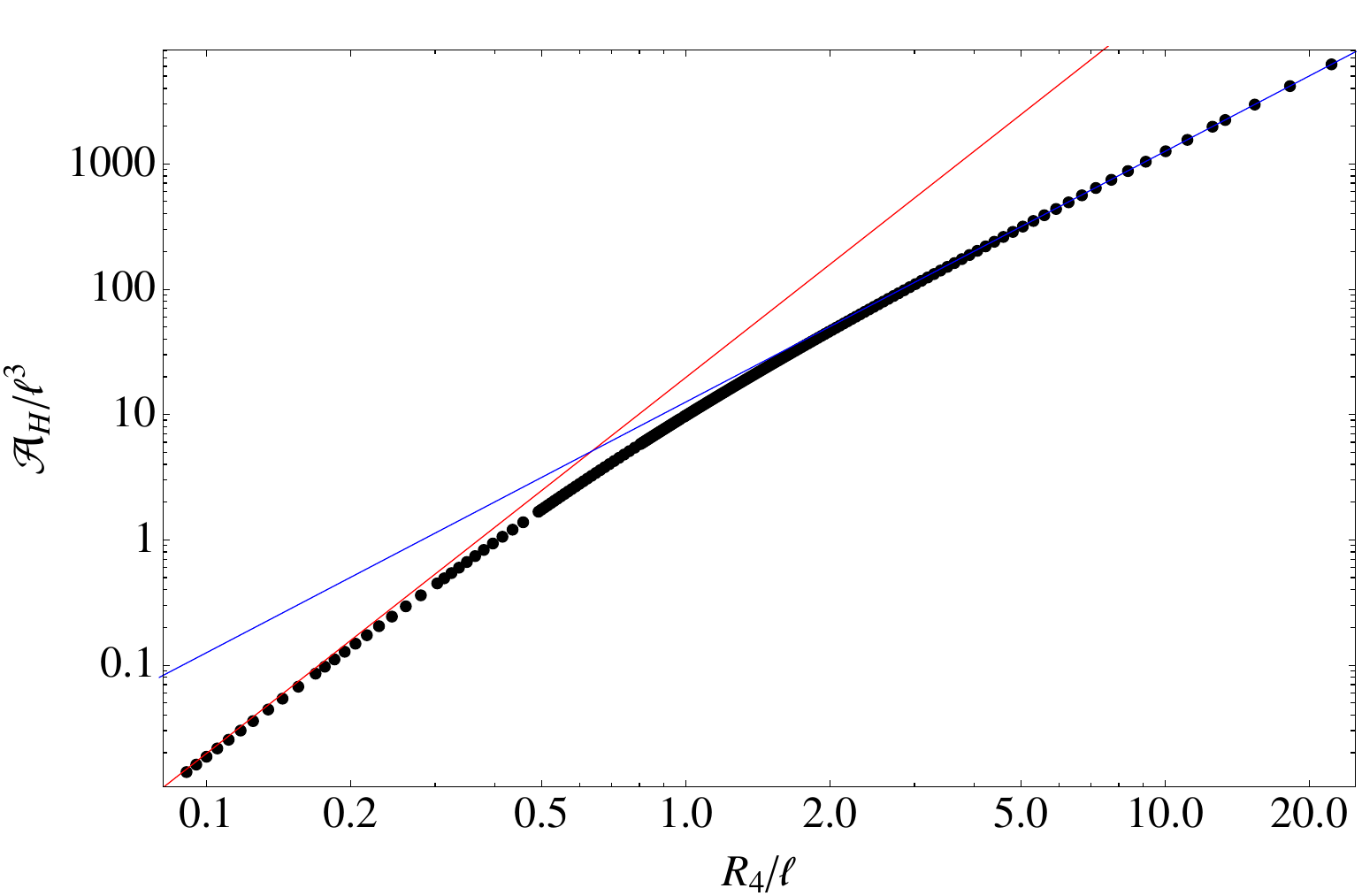}
\end{center}
\caption{Area of the black hole as a function of the radius of the horizon on the brane (black dots), and  the same quantity for an asymptotically flat Schwarzschild black hole in 5$d$ (red) and in 4$d$ (blue). Note the log scale of both axes. }
\label{fig:entropy}
\end{figure}

In Fig. \ref{fig:entropy} we have plotted the area of the full 5$d$ black hole as a function of the radius of the horizon on the brane, comparing it with the analogous quantities for an asymptotically flat Schwarzschild black hole in 5$d$ (red) and 4$d$ (blue) respectively. It is apparent from this plot that small (compared to $\ell$) braneworld black holes behave like 5$d$ asymptotically flat Schwarzschild black holes and large ones recover 4$d$ behaviour.

\begin{figure}[h]
\begin{center}
\includegraphics[scale=0.6]{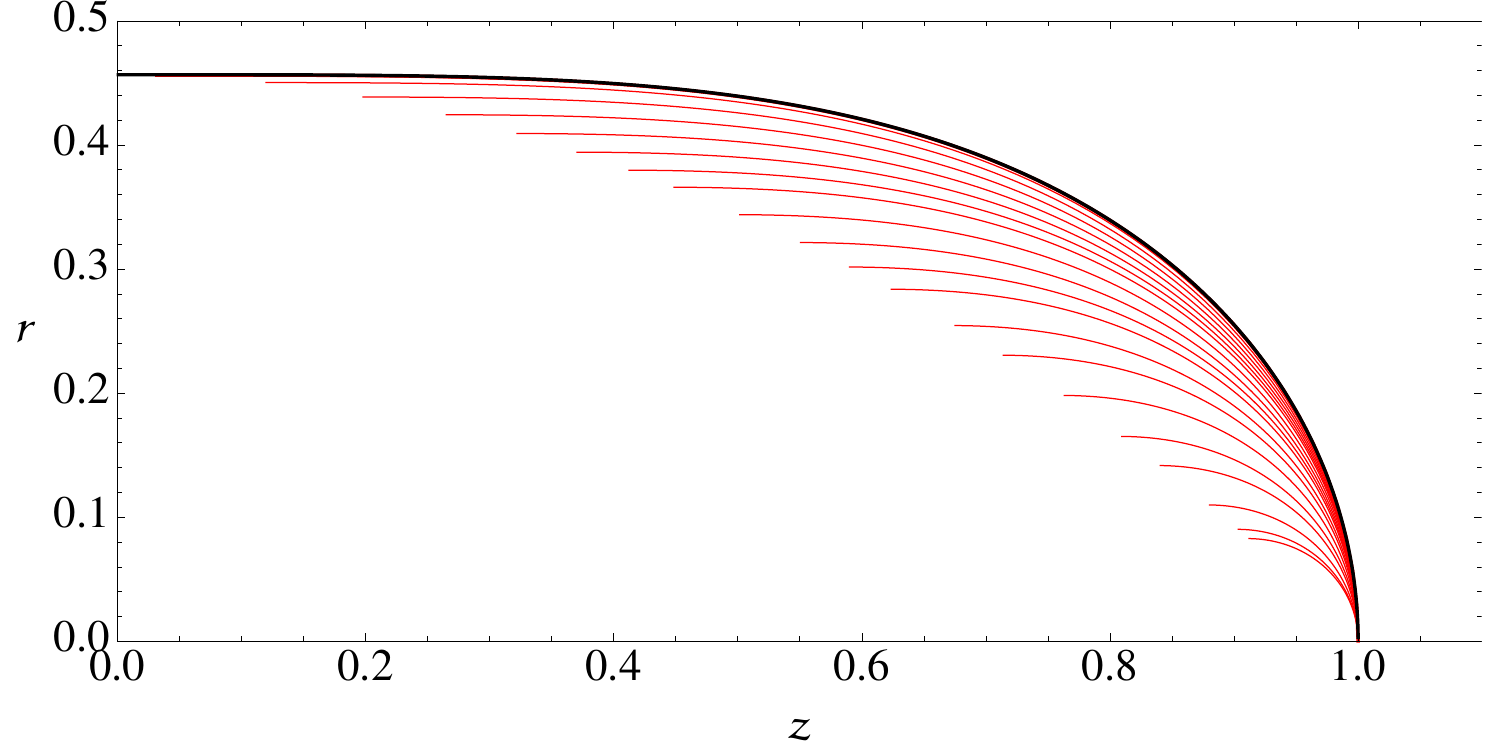}
\end{center}
\caption{Embedding of the spatial cross sections of the  horizon into $\mathbb H^4$ (red). The black curve corresponds to the embedding of the horizon of the $AdS_5$-$CFT_4$ solution of \cite{PFLuciettiTW}, with 4$d$ Schwarzschild as the conformal boundary metric.}
\label{fig:embedding}
\end{figure}

We have embedded  the geometry of the spatial cross sections of the horizon into $\mathbb H^4$, $ds^2=\frac{\ell^2}{z^2}(dz^2+dr^2+r^2d\Omega_{(2)}^2)$, as a surface of revolution $r(z)$ such that the induced metric on this surface is that of the horizon. To compare black holes of different sizes we have fixed the maximum extent of the horizon into the bulk to be at $z=1$ so that the brane is located at a $z=z_{min}$ which depends on the size of the black hole.  Fig. \ref{fig:embedding} depicts the embeddings of the horizon of  braneworld black holes of different sizes (red), together with the embedding of the $AdS_5$-$CFT_4$ solution of \cite{PFLuciettiTW}. This gives a beautiful graphical confirmation of the analysis given in the first part of this letter. For large black holes, where $z_{min} \to 0$, we see the horizon tends to that of the $AdS_5$-$CFT_4$ solution, the perturbation from it getting smaller as the cut off $z_{min}$ is removed. 
From these embeddings we see the horizon is pancake-like \cite{Chamblin:1999by}, with a proper distance to the tip going as $\sim \ell \log(R_4/\ell)$ for large $R_4/\ell$. 
We note the similarity of these large radius embeddings to the shape estimated from linear theory in \cite{Fitzpatrick:2006cd}. As discussed there, such a horizon geometry presumably has too little extent into the bulk to experience a Gregory-Laflamme type instability \cite{Gregory:2000gf}.

We can provide evidence of dynamical stability for our solution by computing the spectrum of our linearized Euclidean Einstein-DeTurck equation about our solutions. We note this linear operator must be computed anyway as part of the Newton method.
For transverse traceless perturbations about an Einstein solution it coincides with the spectrum of the Lichnerowicz operator restricted to static axisymmetric modes \cite{Headrick:2009pv}. We find that for all our solutions there is a single negative mode, which for small $R_4/\ell$ tends to the usual negative mode of 5$d$ 
asymptotically flat 
Schwarzschild black hole. Small solutions are close to 5$d$ Schwarzschild and should be stable. Then the absence of any zero modes and hence new negative modes as one moves to larger radius solutions indicates we should expect no dynamical instabilities, at least with axisymmetry.
Further details will be provided in \cite{PFTW}.

\begin{figure}[h]
\begin{center}
\includegraphics[scale=0.7]{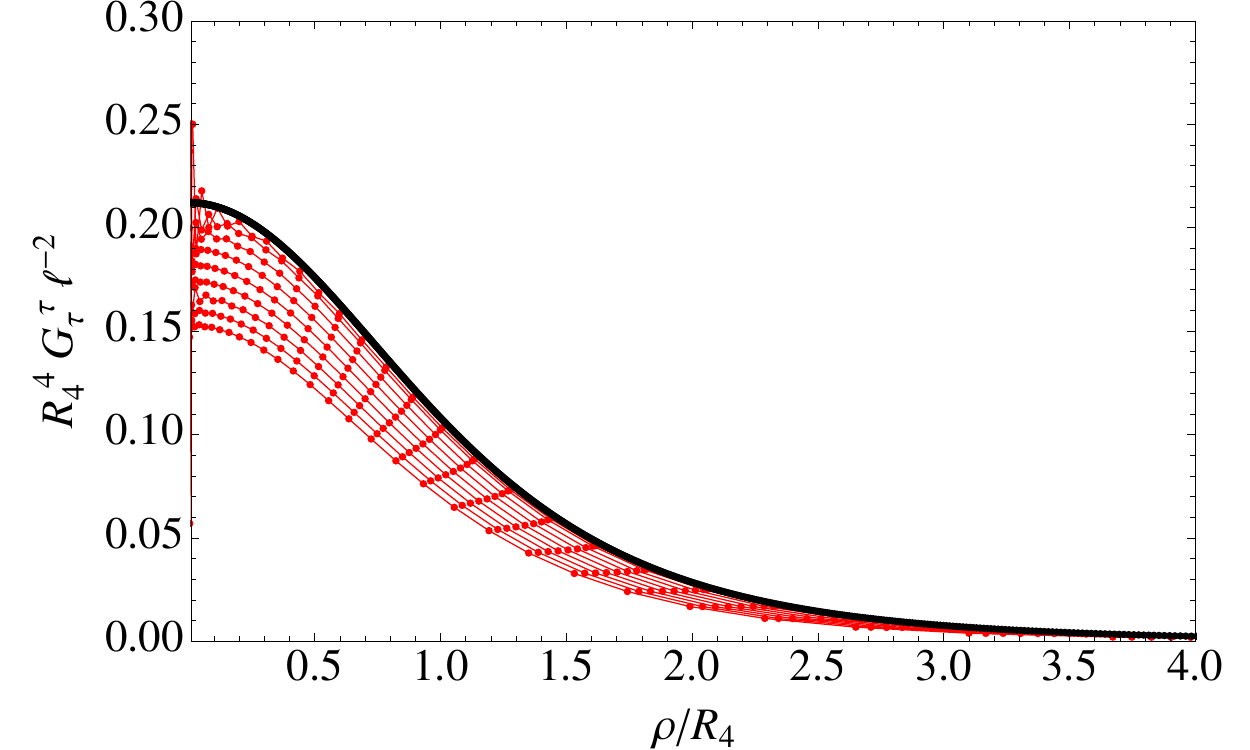}
\end{center}
\caption{ $R_4^{~4} \, G_\tau^{~\tau}\,\ell^{-2}$ computed from the induced geometry on the brane  against proper distance for braneworld black holes of  sizes $R_4/\ell\sim 1.24-6.70$ (red). In black, r.h.s. of \eqref{eq:vac}  computed from the solution of \cite{PFLuciettiTW} using standard holographic renormalisation \cite{deHaro:2000xn}. The red curves approach the black one as the black hole size is increased. For large black holes, the actual value of $G_\tau^{~\tau}$ on the brane is so small as to be comparable to the numerical error, and we see some noise in this quantity. }
\label{fig:Gtt}
\end{figure}

As the black hole becomes larger,  the induced geometry on the brane tends to the 4$d$ asymptotically flat Schwarzschild solution. 
We may verify this by computing the induced Einstein tensor on the brane. In  Fig.  \ref{fig:Gtt} we plot 
 the dimensionless quantity $R_4^{\phantom 44} \, G_\tau^{~\tau}\,\ell^{-2}$   against proper radial distance from the horizon along the brane, 
 $\rho$, in the combination $\rho / R_4$. 
 The other components of $G_{\mu}^{\phantom\mu\nu}$ give the same behaviour, and we see that the solutions become Ricci flat with corrections going as $O(\ell^2/R_4^2)$, i.e. $O(\epsilon^2)$.
With these scalings we see the curves for large radius solutions limit to a fixed curve, which appears to be precisely predicted by the stress tensor of the $AdS_5$-$CFT_4$ solution of \cite{PFLuciettiTW}. This explicitly confirms the prediction \eqref{eq:vac}.   We should note that as the black hole on the brane becomes large the value of $G_\mu^{\phantom\mu\nu}$  becomes small and ultimately, for sufficiently large black holes (while keeping the resolution fixed), e.g. $R_4/\ell\sim 10$,  is comparable to the our numerical error. This is manifest in Fig.  \ref{fig:Gtt}, where  $G_\mu^{\phantom\mu\nu}$ is further multiplied by a  factor of $R_4^{\phantom  4 4}$, which can be very large, and numerical errors contaminate the data for large black holes. 

\begin{figure}[h]
\begin{center}
\includegraphics[scale=0.7]{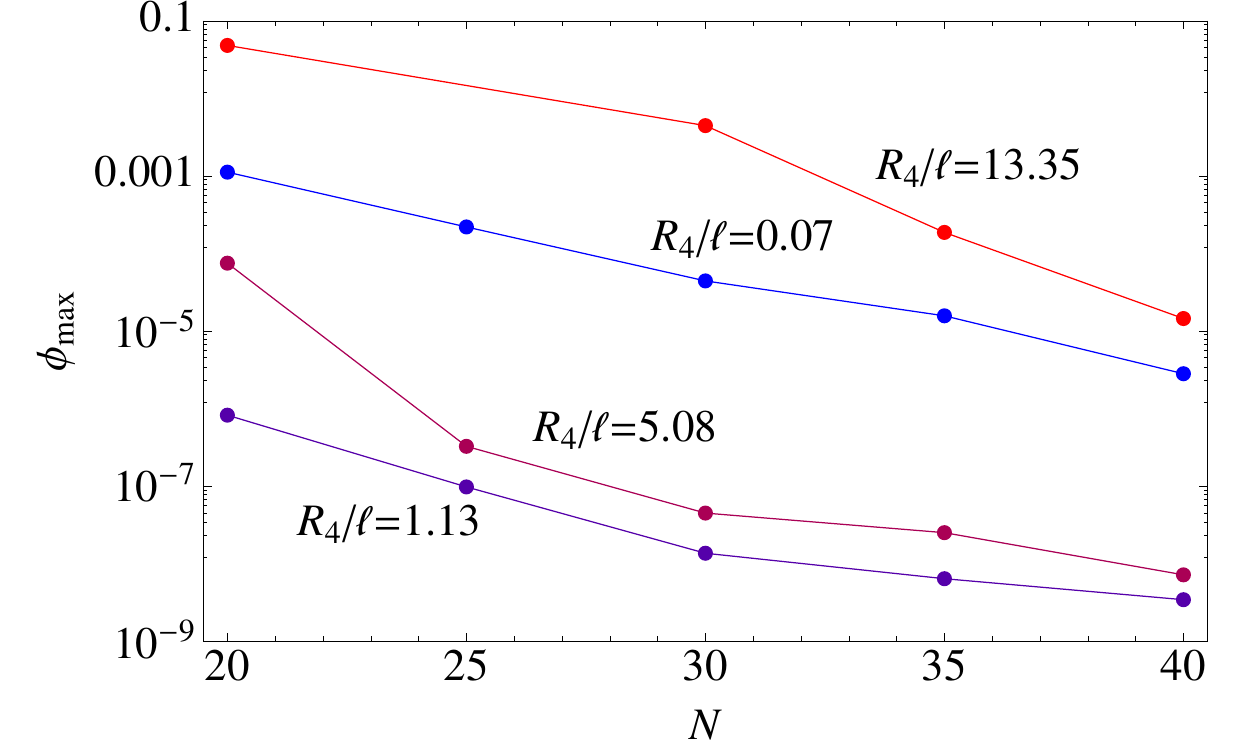}
\end{center}
\caption{Maximum value of $\phi$, $\phi_\textrm{max}$, in the whole domain (including the brane) for braneworld black holes with $R_4/\ell=13.35, 5.04, 1.13, 0.07$, as a function of the number of grid points $N$ for the pseudospectral code.  Lines are drawn to guide the eye.  As this figure shows, $\phi_\textrm{max}\to 0$ in the continuum limit, which provides evidence that our solutions are \textit{not} Ricci solitons. }
\label{fig:convergence}
\end{figure}

Finally we comment on the possibility that our solutions are in fact Ricci solitons. We have performed convergence tests which indicate that the solutions indeed have $\phi = \xi^A \xi_A \rightarrow 0$ in the continuum limit.
As shown in Fig. \ref{fig:convergence} for black holes with $R_4/\ell=O(1)$ our $40 \times 40$  pseudospectral code gives a maximum value of $\phi$, denoted by $\phi_\textrm{max}$, such that $\phi_\textrm{max} < 10^{-8}$, which is already very small. It is also worth noting that  for a fixed spatial resolution, $\phi_\textrm{max}$ grows as the black hole becomes very large or very small, which is expected since we have to resolve widely separated length scales, namely the horizon radius on the brane, and the AdS length.

Another check is to observe that the Bianchi identity for \eqref{eq:DeTurck} implies, $\nabla^2\,\xi^M+R^{M}_{\phantom MN}\xi^N=0$.
Hence a necessary condition for a solution to be a soliton is that the linear elliptic vector operator $D \equiv \delta_A^{~B} \nabla^2 +R_{A}^{~B}$ has a zero mode with 
the same boundary conditions on the vector as for the behaviour of $\xi$ with our boundary conditions.
We have computed the lowest eigenmode of $D$ by simulating long time diffusion $\dot{v}^A = D v^A$ for a vector $v^A$ with the appropriate  boundary behaviour and find no evidence of zero or even near zero modes. This  again confirms that our solutions are not solitons.

We reiterate that the overview of the numerical construction given here will be expanded in detail in the forthcoming work\cite{PFTW}.

\section*{Acknowledgements}

We would like to thank Roberto Emparan, Mukund Rangamani, Harvey Reall and especially James Lucietti for very useful discussions. PF is supported by an EPSRC postdoctoral fellowship [EP/H027106/1]. 
TW is supported by an STFC advanced fellowship and Halliday award.

\bibliographystyle{apsrev}
\bibliography{rsbh}

\end{document}